\documentclass[12pt]{article}

\usepackage{amssymb}
\usepackage{amsmath}
\usepackage{amscd}
\usepackage{latexsym}
\usepackage{graphicx}

\usepackage{cite}

\newcommand{\be}{\begin{equation}}
\newcommand{\ee}{\end{equation}}
\newcommand{\Dlt}{\Delta}
\newcommand{\dlt}{\delta}
\newcommand{\prt}{\partial}
\newcommand{\br}{{\bf r}}

\newcommand{\ba}{{\bf a}}

\newcommand{\bt}{\beta}

\newcommand{\ep}{\varepsilon}
\newcommand{\al}{\alpha}
\newcommand{\ra}{\rightarrow}
\newcommand{\sgm}{\sigma}

\newcommand{\gm}{\gamma}
\newcommand{\om}{\omega}

\newcommand{\Gm}{\Gamma}
\newcommand{\dgr}{\dagger}
\newcommand{\lbd}{\lambda}

\newcommand{\rgl}{\rangle}
\newcommand{\lgl}{\langle}

\begin{document}

\hspace{13cm} {\it Brief review}

\vskip 0.5cm

\begin{center}

{\Large{\bf Bose-Einstein condensation temperature of weakly interacting atoms} \\ [5mm]

V.I. Yukalov$^{1,2}$ and E.P. Yukalova$^3$}  \\ [5mm]

{\it
$^1$Bogolubov Laboratory of Theoretical Physics, \\
Joint Institute for Nuclear Research, Dubna 141980, Russia \\ [3mm]

$^2$Instituto de Fisica de S\~ao Calros, Universidade de S\~ao Paulo, \\
CP 369,  S\~ao Carlos 13560-970, S\~ao Paulo, Brazil \\ [3mm]

$^3$Laboratory of Information Technologies, \\
Joint Institute for Nuclear Research, Dubna 141980, Russia  }

\end{center}

\vskip 5mm

\begin{abstract}
The critical temperature of Bose-Einstein condensation essentially depends on internal
properties of the system as well as on the geometry of a trapping potential. The 
peculiarities of defining the phase transition temperature of Bose-Einstein condensation 
for different systems are reviewed, including homogenous Bose gas, trapped Bose atoms,
and bosons in optical lattices. The method of self-similar approximants, convenient for 
calculating critical temperature, is briefly delineated.      
\end{abstract}

\vskip 3mm
{\parindent =0pt
{\bf Keywords}: Bose-Einstein condensation, critical temperature, homogeneous gas, 
trapped atoms, optical lattices, self-similar approximants}

\vskip 5mm

{\bf Contents}

\vskip 2mm
1. Introduction

\vskip 2mm
2. Homogeneous Bose gas

\vskip 2mm
3. Trapped Bose gas

\vskip 2mm
4. Power-law traps

\vskip 2mm
5. Finite-size corrections

\vskip 2mm
6. Quantum corrections

\vskip 2mm
7. Interaction corrections

\vskip 2mm
8. Box-shaped trap

\vskip 2mm
9. Optical lattices

\vskip 2mm
10. Conclusion

\vskip 2mm
{\bf Appendix}. Self-similar factor approximants

\vskip 2mm
{\bf References}

\newpage

\section{Introduction}

Critical temperature is one of the main characteristics of  systems experiencing
Bose-Einstein condensation (BEC) phase transition. This temperature essentially 
depends on the system parameters and on the geometry of traps confining atomic 
clouds. In the present article, we give a survey of the BEC critical temperatures
for weakly interacting Bose gases in typical systems, such as homogeneous Bose
gas, harmonically trapped Bose gas, bosonic atoms confined by different power-law 
potentials, and Bose gas in an optical lattice. 

We concentrate our attention on the systems for which it is possible to get analytical
expressions for critical temperature. This is because having in hands an analytical,
even maybe approximate, expression for $T_c$ reveals the explicit role of the 
system parameters and makes it clear what are the optimal conditions for realizing
Bose-Einstein condensation.   
  
In the process of calculations, it is sometimes necessary to resort to nontrivial 
theoretical methods. One such an approach, allowing for relatively simple and 
accurate calculations, called self-similar approximation theory, is sketched in the 
Appendix.     

Throughout the paper, we use the system of units, where the Planck and Boltzmann
constants are set to unity, $k_B = 1$ and $\hbar = 1$.

\section{Homogeneous Bose gas}

The ideal homogeneous Bose gas in three dimensions, as is well known, exhibits 
Bose-Einstein condensation at the critical temperature
\be
\label{1}
 T_0 = \frac{2\pi}{m} \left [ \frac{\rho}{\zeta(3/2)} \right ]^{2/3} \;  ,
\ee
where $m$ is atomic mass, $\rho$ is average particle density, and $\zeta(n)$ is 
the Riemann zeta function. The question, attracting for long time attention, is how 
this expression varies under switching on atomic interactions. This problem turned 
out to be highly nontrivial because the ideal Bose gas and the interacting Bose gas, 
even with asymptotically weak interactions, pertain to different classes of universality.
The ideal Bose gas enjoys the Gaussian universality class, while the interacting 
three-dimensional Bose gas pertains to the ${\mathcal O}(2)$, or $XY$, universality 
class. Although the phase transition in both these cases is of continuous second 
order, but the physics in the vicinity of the phase transition is of different nature.
Close to the phase transition, the physics in an interacting gas is governed by strong
fluctuations. As a result, perturbation theory in powers of interaction strength becomes
inapplicable, yielding infrared divergences.  

One usually considers dilute Bose gas, characterized by the local interaction potential 
$$
\Phi(\br) = 4\pi\; \frac{a_s}{m} \; \dlt(\br) \;   ,
$$
in which $a_s$ is the $s$-wave scattering length. One keeps in mind repulsive 
interactions, with positive scattering length $a_s > 0$, since a homogeneous system
with a negative scattering length is unstable \cite{TerHaar_1}. The interaction strength 
of dilute gas is conveniently described by the dimensionless {\it gas parameter}
\be
\label{2}
 \gm \equiv \rho^{1/3} a_s \;  .
\ee
One studies how the critical temperature $T_c$ of an interacting Bose gas shifts
from the temperature $T_0$ of the ideal gas, when switching on atomic interactions.
The relative temperature shift is defined as
\be
\label{3}
 \frac{\Dlt T_c}{T_0} \equiv \frac{T_c-T_0}{T_0} \;  .
\ee

For an asymptotically small gas parameter $\gamma \ra 0$, the critical temperature 
shift behaves as \cite{Holzmann_2,Arnold_3}
\be
\label{4}
\frac{\Dlt T_c}{T_0} \simeq c_1\gm + ( c_2 + c_2'\ln \gm) \gm^2 \; .
\ee
The coefficient $c_2'$ can be found exactly using perturbation theory \cite{Arnold_3}
that gives 
\be
\label{5}
 c_2' = - \; \frac{64\pi\zeta(1/2)}{3[\zeta(3/2)]^{5/3}} = 19.7518 \;  .
\ee
While for $c_1$ and $c_2$ perturbation theory fails, and one needs more elaborate
calculational methods.

There have been numerous attempts of calculating the nonperturbative coefficients
$c_1$ and $c_2$, employing different techniques, such as Ursell operators and Green 
functions, renormalization group, the $1/N$-expansion in the $N$-component field 
theory, and so on, as summarized in the review articles \cite{Andersen_4,Yukalov_5}. 
The results ranged in wide intervals. Thus for $c_1$ one obtained  the values between
$-0.95$ and $4.7$ and for $c_2$ the values ranging from $4.9$ to $101.4$, as 
discussed in Refs. \cite{Andersen_4,Yukalov_5}. Numerical calculations, using
Monte Carlo simulations for three-dimensional lattice ${\mathcal O}(2)$ field theory, 
give $c_1 = 1.29 \pm 0.05$ \cite{Kashurnikov_6,Prokofev_7} and $c_1 = 1.32 \pm 0.02$ 
\cite{Arnold_8,Arnold_9}, while $c_2 =  75.7\pm 0.4$  \cite{Arnold_3}.

{\it Optimized perturbation theory}, advanced in Refs. \cite{Yukalov_10,Yukalov_11}, 
has also been used for calculating the coefficients $c_1$ and $c_2$. The main idea 
of optimized perturbation theory is to define control functions making asymptotic 
perturbative sequences convergent. Control functions can be introduced in three ways: 
either by including them into an initial approximation, e.g., into the initial Lagrangian or 
Hamiltonian, or incorporating them into a sequence transformation, or  including them 
into a change of variables \cite{Yukalov_12}. Optimized perturbation theory has been 
used for calculating  the coefficients $c_1$ and $c_2$ by two methods: including 
control functions into an initial approximation of Lagrangian 
\cite{Pinto_13,Souza_14,Souza_15,Souza_16,Braaten_17,Braaten_18,Kneur_19,
Kneur_20,Kneur_21,Farias_22} and by introducing them through the Kleinert 
\cite{Kleinert_23} change of variables \cite{Kastening_24,Kastening_25,Kastening_26} . 
Both ways give results close to those of  Monte Carlo simulations.    

Determining the condensation temperature of a Bose gas can be reformulated as the 
problem of defining the critical temperature for a three-dimensional $N$-component field 
theory \cite{Arnold_3,Kastening_25,Kastening_26}. Different $N$ correspond to 
different physical systems. Thus $N = 0$ corresponds to dilute polymer solutions, $N = 1$,
to the Ising model, $N = 2$, to magnetic $XY$ models and to superfluids, and $N = 3$
characterizes the Heisenberg model. 

When perturbation theory is straightforwardly applied to the calculation of the coefficients  
$c_1$ and $c_2$, the following problem arises. Loop expansion yields asymptotic series 
in powers of the variable
$$
 x = \frac{\lbd_{eff}}{\sqrt{\mu_{eff}}} \; (N+2) \;  ,
$$
in which $N$ is the number of components, $\lambda_{eff}$ is an effective coupling 
parameter, and $\mu_{eff}$ is an effective chemical potential \cite{Kastening_26}.
Then the coefficient, say $c_1$, is represented as an asymptotic series in powers 
of this variable,
\be
\label{6}
 c_1(x) \simeq \sum_n a_n x^n \qquad (x\ra 0) \;  .
\ee
In the case of  the seven-loop expansion \cite{Kastening_26}, the coefficients $a_n$ 
for different numbers of the field components $N$ are listed in Table 1.

However at $T_c$, the effective chemical potential tends to zero, $\mu_{eff} \ra 0$,  
because of which the variable $x$ tends to infinity, $x \ra \infty$. Hence formally we 
need to find the limit
\be
\label{7}
 c_1 = \lim_{x\ra\infty} c_1(x) \;  .
\ee
Of course, this limit has no sense being applied directly to series (\ref{6}). 
Before taking the limit, it is necessary to extrapolate the series for the 
asymptotically small variable to its arbitrary values, including asymptotically 
large values. Such an extrapolation is provided by optimized perturbation theory, 
as has been done in Refs. 
\cite{Pinto_13,Souza_14,Souza_15,Souza_16,Braaten_17,Braaten_18,Kneur_19,
Kneur_20,Kneur_21,Farias_22,Kastening_24,Kastening_25,Kastening_26}.

Another method of extrapolation is based on {\it self-similar approximation theory}
\cite{Yukalov_27,Yukalov_28,Yukalov_29,Yukalov_30,Yukalov_31,Yukalov_32,Yukalov_33}.
Employing the extrapolation by {\it self-similar factor approximants} 
\cite{Yukalov_34,Gluzman_35,Yukalov_36}, as described in the Appendix, we find 
\cite{Yukalov_73} the values of $c_1$ summarized in Table 2. As is seen, these 
values are very close to the available Monte Carlo simulations for $N = 2$ 
\cite{Kashurnikov_6,Prokofev_7,Arnold_8,Arnold_9} and for $N = 1$ and $N = 4$ 
\cite{Sun_37}.

Note that for the formal limit $N \ra \infty$, the coefficient $c_1$ is known exactly 
\cite{Baym_38}, being
$$
\lim_{N\ra\infty} c_1 = \frac{8\pi}{3[\zeta(3/2)]^{4/3} } = 2.328473 \; .
$$
 
For the coefficient $c_2$, one finds the expression that we write here in a slightly 
different notation, as compared to that of Kastening \cite{Kastening_26},
$$
c_2 =  \frac{32\pi\zeta(1/2)}{3[\zeta(3/2)]^{5/3} }\;
\left \{ \ln \;\frac{[\zeta(3/2)]^{2/3}}{128\pi^3} + 
\frac{\ln 2}{\sqrt{\pi}} \; \zeta \left( \frac{1}{2} \right ) +
\frac{\sqrt{\pi}+0.270166}{\zeta(1/2)} \; - \; 1 \right \} +
$$
\be
\label{8}
+ \frac{7}{4} \; c_1^2 + \frac{8\zeta(1/2)}{[\zeta(3/2)]^{1/3} }\;
\left ( 3d_2 - 2c_1 \right ) \; .
\ee
In the seven-loop expansion, one has \cite{Kastening_26}
\be
\label{9}
d_N(x) = \frac{1}{x^2} \left ( \sum_{n=0}^7 b_n x^n + b_2' x^2 \ln x \right ) \; .
\ee
Several first coefficients $b_n$ can be written explicitly as
$$
b_0 = (N+2) A_0 \; , \qquad b_1 = \frac{N+2}{24\pi} \; A_0 \; ,
$$
$$
b_2 = \frac{1-4\ln 6}{576\pi^2} \; A_0 = - 2.393297 \; , \qquad
b_2' = \frac{A_0}{288\pi^2} = 0.776158 \; ,
$$
where 
$$
 A_0 = \frac{256\pi^3}{[\zeta(3/2)]^{4/3} } = 2206.18611757 \;  .
$$
These and the higher-order coefficients $b_n$ are given in Table 3. 

Again, one needs to extrapolate the asymptotic series (\ref{9}) to finite values of the
variable $x$ and to find an effective limit
\be
\label{10}
d_N = \lim_{x\ra\infty} d_N(x) \;   .
\ee
In Table 4, the results for $d_N$, obtained by Kastening  \cite{Kastening_26}, 
are presented, compared with the available Monte Carlo simulations for $N = 2$ 
\cite{Arnold_9} and for $N = 1, 4$ \cite{Sun_37}. Thus for Bose-Einstein condensation
with $N = 2$, the Monte Carlo simulations give $c_2 = 75.7 \pm 0.4$ \cite{Arnold_9}.   

We may note that the large $N$ limit for $d_N$ is known exactly \cite{Kastening_26},
being
$$
\lim_{N\ra\infty} d_N = \frac{1+2\ln 2}{288\pi^2}\; A_0 = 1.8521396 \; .
$$

By Monte Carlo simulations, one can find the critical temperature not only for weak 
interactions, when the gas parameter is small, $\gamma \ra 0$, but also for finite $\gamma$,
as has been done by Pilati et al. \cite{Pilati_39} for  $N = 2$, whose results are presented 
in Table 5. In Fig. 1, we show the relative critical temperature of Bose-Einstein condensation
$T_c/T_0$ as a function of the gas parameter $\gamma$, following from Eq. (\ref{4}), as 
compared with the Monte Carlo results from Table 5. As is clear, Eq. (\ref{4}) is valid only 
for $\gamma \ra 0$, as one could expect.  In the simulations of Pilati et al. \cite{Pilati_39},
a Bose gas of hard spheres is considered, where $a_s$ corresponds to the hard-sphere 
diameter. The hard-sphere system is often used as a reference system for realistic fluids,
such as liquid helium \cite{Kalos_40,Solis_41}. When $a_s$ is much shorter than the mean
interatomic distance $a$, the results for the local pseudopotential coincide with those 
for the hard-sphere system \cite{Giorgini_42}.  Moreover, the results for the local potential 
can be extended \cite{Yukalov_43} to finite values of the ratio $a_s/a$ up to $a_s/a =0.65$, 
where the fluid freezes \cite{Rossi_44}.  Generally, it is possible to show that 
pseudopotentials can be employed for a rather accurate modeling of physical systems, 
including those with nonintegrable interaction potentials \cite{Yukalov_45}. In the case 
of superfluid $^4$He at saturated vapor pressure, the effective hard-sphere diameter is 
$a_s = 2.203 \AA$, which corresponds to the gas parameter $\gamma = 0.5944$
and to the relative critical temperature $T_c/T_0 = 0.7$. 

Notice that the critical temperature, at asymptotically weak interactions increases
with $\gamma$. This is because an ideal homogeneous Bose-condensed gas is 
unstable, while interactions stabilize it \cite{Yukalov_46,Yukalov_47}. At the same time, 
strong interactions destroy the condensate, because of which at higher values of 
$\gamma$ the critical temperature diminishes.

\section{Trapped Bose gas}

Many finite quantum systems are well represented as being confined by effective
potentials \cite{Birman_90}. Most often, one considers a three-dimensional harmonic 
trapping potential 
\be
\label{11}
 U(\br) = \frac{m}{2} \left ( \om_x^2 x^2 + \om_y^2 y^2 + \om_z^2 z^2 \right ) \;  .
\ee
Strictly speaking, genuine Bose-Einstein condensation happens only in an infinite 
system. For a gas confined by the harmonic potential, this implies \cite{Pitaevskii_48} 
the limits
\be
\label{12}
N \ra \infty \; , \qquad \om_0 \ra 0 \; , \qquad N\om_0^3 \ra const \; .
\ee
Here $N$ is the number of atoms in the trap and 
\be
\label{13}
 \om_0 \equiv ( \om_x \om_y \om_z )^{1/3} \;  .
\ee
For finite $N$ there occurs pseudocondensation or quasicondensation \cite{Mullin_49},
which in what follows, for short, will also be called condensation.   

The ideal Bose gas in a harmonic trap condenses at the critical temperature    
\be
\label{14}
 T_0 = \left [ \frac{N}{\zeta(3) } \right ]^{1/3} \om_0 \;  .
\ee
At this temperature, the thermal wavelength is
\be
\label{15}
\lbd_0 \equiv \sqrt{\frac{2\pi}{mT_0} } = \sqrt{\frac{2\pi}{m\om_0} } 
\left [ \frac{\zeta(3)}{N} \right ]^{1/6} \;  .
\ee
The ratio
\be
\label{16}
\al \equiv \frac{a_s}{\lbd_0}
\ee
plays the role of a dimensionless coupling parameter. The critical temperature shift
in terms of the asymptotically small parameter reads as \cite{Arnold_50}
\be
\label{17}
 \frac{\Dlt T_c}{T_0} \simeq \overline c_1 \al + \left ( \overline c_2 +
\overline c_2' \ln \al \right ) \al^2 \;  .
\ee
The first coefficient is known \cite{Pitaevskii_48,Mullin_49} from perturbative 
calculations, and the coefficient $\bar{c}_2'$ can also be calculated  perturbatively 
\cite{Arnold_50},
\be
\label{18}
\overline c_1 = - 3.426032 \; , \qquad  
\overline c_2' = - \; \frac{32\pi\zeta(2)}{3\zeta(3)} = - 45.856623 \; .
\ee
The coefficient $\bar{c}_2$ can be related to lattice simulations in three-dimensional
${\mathcal O}(2)$ field theory  \cite{Arnold_50}, 
\be
\label{19}
 \overline c_2 = 21.4 - \; \frac{16\pi\zeta(2)}{3\zeta(3)} \; \left [
\ln (32\pi^3 ) + 3.522272 \right ] + 
12\zeta(2) \left [ \zeta \left (\frac{3}{2} \right ) \right ]^{4/3} d_2 \;  ,
\ee
with the same $d_2$ as in Eq. (\ref{9}).  Using the Monte Carlo result \cite{Arnold_9}
for $d_2$, one has 
$$
 \overline c_2 = - 155.0 \;  .
$$

We may notice the difference with the case of a homogeneous gas, where the linear 
term is positive, while for the trapped gas, according to Eq. (\ref{18}), it is negative. 
This is because the ideal Bose-condensed gas in a three-dimensional harmonic trap 
is stable, so that switching on interactions immediately starts 
destroying the condensate. While, on the contrary, the ideal homogeneous 
Bose-condensed gas is unstable. Therefore interactions play the dual role, first 
stabilizing the system and then, when the interactions become sufficiently strong, 
they start depleting the condensate \cite{Yukalov_5,Yukalov_47}.

The critical temperature shift due to repulsive interactions was studied experimentally
for harmonically trapped $^{87}$Rb atoms \cite{Gerbier_51} and $^{39}$K atoms 
\cite{Smith_52,Smith_53}, varying the interaction strength by means of Feshbach 
resonance. The measurements were found to be in good agreement with the first 
coefficient $\bar{c}_1 \approx -3.426$. However the next term was different, as 
compared with Eq. (\ref{17}). Exploring the range $0.001 < \alpha < 0.06$, all 
experimental data points were fitted \cite{Smith_52,Smith_53} by the second-order 
polynomial
\be
\label{20}
\frac{\Dlt T_c}{T_0} = b_1 \al + b_2 \al^2 \;   ,
\ee
with the coefficients 
$$
 b_1 = \overline c_1 = - 3.426 \; , \qquad b_2 = 46 \pm 5 \;  .
$$
The second-order positive term is due to atomic correlations beyond mean-field picture
\cite{Houbiers_54,Holzmann_55,Zobay_56,Briscese_57,Castellanos_58}.
In order to show how expressions Eq. (\ref{17}) and Eq. (\ref{20}) are different, they are 
presented in Fig. 2.

\section{Power-law traps}

Traps, confining atoms, can be not only harmonic, but more generally, of power law, having
in $d$-dimensional space the form 
\be
\label{21}
U(\br) = \sum_{\al=1}^d \frac{\om_\al}{2} \left | \frac{r_\al}{l_\al} \right |^{n_\al}
\qquad \left ( l_\al \equiv \frac{1}{\sqrt{m\om_\al} } \right ) \; .
\ee
The characteristic trap frequency and length are defined as
\be
\label{22}
\om_0 \equiv \left ( \prod_{\al=1}^d \om_\al \right )^{1/d} = \frac{1}{ml_0^2} \; , \qquad
l_0 \equiv \left ( \prod_{\al=1}^d l_\al \right )^{1/d} = \frac{1}{\sqrt{m\om_0} }   \; ,
\ee
respectively. For what follows, it is convenient to introduce the {\it confining dimension}
\cite{Yukalov_47,Yukalov_59}
\be
\label{23}
 s \equiv \frac{d}{2} + \sum_{\al=1}^d \frac{1}{n_\al} \; .
\ee

Let us consider the ideal Bose gas trapped inside potential (\ref{21}). In the 
semiclassical description, we have the density of states
\be
\label{24}
 \rho(\ep) = \frac{\ep^{s-1}}{\gm_d\Gm(s) } \qquad 
\left ( s \geq \frac{1}{2} \right ) \;  ,
\ee
in which
\be
\label{25}
 \gm_d \equiv \frac{\pi^{d/2}}{2^s} 
\prod_{\al=1}^d \frac{\om_\al^{1/2+1/n_\al}}{\Gm(1+1/n_\al)} 
\qquad ( d \geq 1) \;  .
\ee

In the standard semiclassical approximation, the energy variable $\varepsilon$
varies between zero and infinity, as a result of which Bose condensation
becomes impossible in some low-dimensional traps 
\cite{Bagnato_60,Bagnato_61,Courteille_62}. However, in quantum systems the
energy varies not from zero, but from a finite value $\varepsilon_0$ corresponding 
to the lowest energy level of the quantum system. By the order of magnitude, it is
possible to approximate the minimal energy as
$$
\ep_0 \approx \frac{k_{min}^2}{2m} \approx \frac{1}{2ml_0^2} = 
\frac{\om_0}{2} \; .
$$
Therefore the range of variation of $\varepsilon$ has to be
\be
\label{26}
\ep_0 \approx \frac{\om_0}{2} \leq \ep < \infty \; .
\ee
Taking this into account defines the {\it modified semiclassical approximation} 
\cite{Yukalov_59}. Then the Bose-Einstein function is modified to the form
\be
\label{27}
g_n(z) = \frac{1}{\Gm(n)} \int_{u_{min}}^\infty \frac{zu^{n-1}}{e^u-z} \; du \; ,
\ee
where $z = e^{\beta \mu}$ is fugacity and the integration is not from zero, but 
from the finite lower limit 
$$
u_{min} \equiv \frac{\om_0}{2T} \; .
$$
The value $\omega_0 / T_0$, where $T_0$ is the critical temperature of trapped 
ideal gas, is assumed to be small. 

Bose-Einstein condensation of ideal gas in a power-law trap occurs (see details in 
Refs. \cite{Yukalov_47,Yukalov_59}) at the temperature
\be
\label{28}
 T_0 = \left [ \frac{\gm_d N}{g_s(1)} \right ]^{1/s} \; .
\ee
In particular, the condensation temperature formally exists in one-dimensional space,
where for an anharmonic trap one has
\be
\label{29}
 T_0 = \frac{\sqrt{\pi} (1-s)\Gm(s)}{2\Gm(1+1/n)} \; N\om_0 \qquad
(d=1 , \; s < 1 , \; n > 2 ) \; .
\ee
For harmonic confinement, we come to the expression that coincides with that
obtained in purely quantum consideration \cite{Ketterle_63},
\be
\label{30}
T_0 = \frac{N\om_0}{\ln(2N)}  \qquad
(s = d =  1 , \; n = 2 ) \;   .
\ee
And for two- and three-dimensional harmonic traps, we have 
\be
\label{31}
T_0 = \om_0 \left [ \frac{N}{\zeta(d)}  \right ]^{1/d} \qquad 
(s = d\geq  2 , \; n_\al = 2 ) \; .
\ee

Recall that Bose-Einstein condensation, strictly speaking, assumes thermodynamic 
limit, which requires to study the system properties under a large number of atoms. 
Usually, in traps this number is really large. In the presence of confining potentials, 
thermodynamic limit is defined differently as compared to homogeneous systems. 

The most general definition of thermodynamic limit, valid for any system, is as follows 
\cite{Yukalov_46,Yukalov_47}: Extensive observable quantities $A_N$ must increase 
together with the number of particles $N$, so that   
\be
\label{32}
N \ra \infty \; , \qquad A_N \ra \infty \; , \qquad \frac{A_N}{N} \ra const \;  .
\ee
As an observable, it is possible to take the system energy $E_N$, so that
\be
\label{33}
N \ra \infty \; , \qquad E_N \ra \infty \; , \qquad \frac{E_N}{N} \ra const \; .
\ee
For the considered case, the latter reduces to the limit
$$
N \ra \infty \; , \qquad \gm_d \ra 0 \; , \qquad N \gm_d \ra const \;   .
$$
In particular, for unipower potentials, when $n_\alpha = n$, we have
\be
\label{34}
N \ra \infty \; , \qquad \om_0 \ra 0 \; , \qquad N\om_0^s \ra const \;   .
\ee

In this limit, the critical temperature (\ref{28}), depending on the confining dimension $s$, 
behaves as
$$
T_0 \; \propto \; \frac{1}{N^{1/s-1}} \ra 0 \qquad ( s < 1) \; ,
$$
$$
T_0 \; \propto \; \frac{1}{\ln N } \ra 0 \qquad ( s = 1) \; ,
$$
\be
\label{35}
T_0 \; \propto \; const \qquad ( s > 1) \; .
\ee
This tells us that a well defined condensation temperature exists only for $s > 1$.

Moreover, a critical temperature may formally exist, but the system below this temperature
looses its stability, which implies that, actually, such a system cannot exist. To check the
stability of a Bose-condensed system, we need to study its compressibility
\be
\label{36}
\kappa_T = \frac{{\rm var}(\hat N)}{\rho T N }  = \frac{1}{\rho^2}
\left ( \frac{\prt\rho}{\prt\mu} \right ) = 
\frac{1}{\rho N} \left ( \frac{\prt N}{\prt\mu} \right ) \;   ,
\ee
in which $\hat{N}$ is the number-of-particle operator and 
$$
 {\rm var}(\hat N) \equiv \lgl \hat N^2 \rgl - \lgl \hat N \rgl^2 \; .
$$
An equilibrium system is stable provided that its particle fluctuations are 
thermodynamically normal \cite{Yukalov_46,Yukalov_64,Yukalov_65}, such that
\be
\label{37}
 0 \leq \kappa_T < \infty \;  .
\ee
In other words, the stability condition is
\be
\label{38}
 0 \leq \frac{{\rm var}(\hat N)}{N} < \infty  \; .
\ee
    
Calculating the compressibility, we find that, above the condensation phase transition,
systems with the confining dimension $s < 1$ are unstable, since
\be
\label{39}
\kappa_T \; \propto \; - N^{1/s-1} \ra - \infty \qquad ( s < 1 , \; T > T_0 ) \;  .
\ee
While for larger confining dimensions, systems are stable, 
\be
\label{40}
 \kappa_T \; \propto \; const \qquad ( s \geq 1 , \; T > T_0 ) \;  .
\ee

The situation is even more delicate below the transition temperature $T_0$, where
the systems up to $s = 2$ are unstable, with 
$$
\kappa_T \; \propto \; - N^{2/s-1} \ra - \infty \qquad ( s < 1 ) \; ,
$$
$$
\kappa_T \; \propto \;  N \ra  \infty \qquad ( s = 1 ) \; , 
$$
$$
\kappa_T \; \propto \;  N^{2/s-1} \ra  \infty \qquad ( 1 < s < 2 ) \; ,
$$
\be
\label{41}
\kappa_T \; \propto \; \ln N  \ra  \infty \qquad ( s = 2 ) \; .
\ee
And only for larger $s > 2$, they are stable, because
\be
\label{42}
\kappa_T \; \propto \; const  \qquad  ( s > 2 ) \;   .
\ee
Remembering the definition of the confining dimension (\ref{23}), we find the stability
condition
\be
\label{43}
 \frac{d}{2} + \sum_{\al=1}^d \frac{1}{n_\al} > 2 \;  .
\ee

For harmonic confinement, when $n_\alpha = 2$, we have
$$
s = d \qquad ( n_\al = 2 ) \;   .
$$
Then we see that Bose-condensed systems in one- and two-dimensional harmonic 
traps are unstable, displaying thermodynamically anomalous particle fluctuations,
$$
\kappa_T \; \propto \; N \ra \infty \qquad ( s = d = 1 ) \; , 
$$
\be
\label{44}
\kappa_T \; \propto \; \ln N \ra \infty \qquad ( s = d = 2 ) \;   .
\ee
And only a three-dimensional harmonic trap can confine a stable Bose-condensed 
system, with a finite compressibility,
\be
\label{45}
\kappa_T \; \propto \; const \qquad ( s = d = 3 ) \;  .
\ee

Although Bose-condensed gas is unstable in one- and two-dimensional harmonic traps,
it can be stable for other powers of the confining potential. As an example, let us consider
the unipower potentials, when $n_\alpha = n$. Then the confining dimension is
$$
 s = \left ( \frac{1}{2} + \frac{1}{n} \right ) d \;  .
$$
According to condition (\ref{43}), the ideal Bose-condensed  gas is stable if the 
confining dimension $s > 2$. Therefore, the gas is stable in different real-space 
dimensions $d$, provided that the potential powers $n$ are limited from above: 
$$
 n < \frac{2d}{4-d} \;  .
$$
That is,
$$
n < \frac{2}{3} \qquad ( d = 1 ) \; ,
$$
$$
n < 2 \qquad ( d = 2 ) \; ,
$$
$$
n < 6 \qquad ( d = 3 ) \; .
$$
The existence of the upper potential power $n$ for stability can be understood, if
one remembers that the passage from a trapped gas to the uniform gas confined in 
a box of length $L$ corresponds to the limits
$$
n \ra \infty \; , \qquad l_0 \ra \frac{L}{2} \; , \qquad s \ra \frac{d}{2} \; .
$$
Hence increasing $n$ approaches the system to the uniform case that is known 
to be unstable \cite{Yukalov_46,Yukalov_47}.

Generally, for stability of a system, it is also necessary that thermal fluctuations 
be thermodynamically normal \cite{Yukalov_46}. Calculating specific heat, we find
\cite{Yukalov_47,Yukalov_59} that it is positive and finite for all $s \geq 1/2$. From 
the definition of the confining dimension $s$ it follows that it is always not smaller 
than $1/2$. Hence thermal fluctuations are thermodynamically normal for the 
considered trapped systems. And stability is defined by the behavior of the 
isothermal compressibility.

\section{Finite-size corrections}

Describing Bose gas in power-law traps, we have used the modified semiclassical 
approximation, accepting the modified function
\be
\label{46}
 g_s(z) = \frac{1}{\Gm(s)} \int_{\bt\ep_0}^\infty \; 
\frac{z u ^{s-1}}{e^u-z} \; du \;  ,
\ee
with a finite lower integration limit, instead of the usual Bose-Einstein function
\be  
\label{47}
 g_s^{(0)}(z) = \frac{1}{\Gm(s)} \int_{0}^\infty \frac{z u ^{s-1}}{e^u-z} \; du \;  ,
\ee
with the zero lower limit.  In the modified variant, the transition temperature
is given by Eq. (\ref{28}), while in the standard case, taking into account that 
$g^{(0)}_s(1) = \zeta(s)$, it is
\be
\label{48}
T_0^{(0)} = \left [ \frac{\gm_d N}{\zeta(s)} \right ]^{1/s} \qquad ( s > 1 ) \; .
\ee
A finite positive value for the latter temperature exists only for $s > 1$, since the 
Rieman zeta function $\zeta(s)$ is finite and positive for $s > 1$, finite but negative 
for $0 \leq s < 1$, and infinite at $s=1$, where $\zeta(s)\ra +\infty$, as $s\ra 1+0$.

In order to study how the use of the modified function (\ref{46}) influences the 
value of the transition temperature, let us consider the critical  temperature shift
under a large number of atoms,
$$
 \frac{\Dlt T_0}{ T_0^{(0)}} \equiv \frac{T_0-T_0^{(0)}}{T_0^{(0)}} \; ,
$$
where $s > 1$, since finite temperature (\ref{48}) exists only for this confining dimension. 
We find \cite{Yukalov_47}
\be
\label{49}
 \frac{\Dlt T_0}{ T_0^{(0)}} = \frac{1}{s(s-1)\Gm(s)\zeta(s) } 
\left ( \frac{\ep_0}{T_0} \right )^{s-1} \;  ,
\ee
which shows that at large $N$ one has
$$
 \frac{\Dlt T_0}{ T_0^{(0)}} \; \propto \; \frac{1}{N^{1-1/s}} \qquad 
( s > 1 ) \; .
$$
For harmonic traps, it follows
$$
\frac{\Dlt T_0}{ T_0^{(0)}} = \frac{1}{4\sqrt{\zeta(2)}\; N^{1/2} } \qquad 
( s=d=2) \; ,
$$
\be
\label{50}
\frac{\Dlt T_0}{ T_0^{(0)}} = \frac{1}{48[\zeta(3)]^{1/3}\; N^{2/3} } \qquad 
( s=d=3) \;   .
\ee

The found critical temperature shifts represent finite-size corrections related to the
quantum nature of trapped atomic systems. A finite quantum system possesses an
energy spectrum with a nonzero energy $\varepsilon_0$ of the lowest level. This has
been taken into account in modifying function (\ref{47})  to (\ref{46}). However, the 
quantum nature of trapped systems also plays the role in a different, and even more
important,  way, as is shown in the following section.

\section{Quantum corrections}

As far as a finite quantum system possesses a spectrum with a nonzero lowest 
energy level $\varepsilon_0$, then at the Bose-Einstein condensation temperature
$T_c$, the chemical potential tends to $\varepsilon_0$, but not to zero. Therefore
the fugacity $z = \exp(\beta \mu)$ tends to $\exp(\varepsilon_0 / T_c)$, but not to one.   
 
In the previous section, we considered how the critical temperature varies when
$g_s^{(0)}(1)$ is replaced by $g_s(1)$. Now, our aim is to study the variation of
the critical temperature under the replacement of  $g_s^{(0)}(1)$ by 
$g_s^{(0)}(e^{\beta \varepsilon_0})$. Similarly to the previous section, we consider 
the confining dimensions $s > 1$. 

Let us denote by $T_c$ the critical temperature calculated with the use of 
$g_s^{(0)}(e^{\beta \varepsilon_0})$ and by $T_0$ the condensation temperature
corresponding to $g_s^{(0)}(1)$. For the relative critical temperature shift, we
find \cite{Yukalov_47} 
$$
\frac{\Dlt T_c}{ T_0} = -\; \frac{(s-1)\ep_0}{s(2-s)\Gm(s)\zeta(s)T_0 } 
\left ( \frac{T_0}{\ep_0} \right )^{2-s} \qquad ( 1 < s < 2) \; ,
$$
$$
\frac{\Dlt T_c}{ T_0} = -\; \frac{\ep_0}{2\zeta(2)T_0} \; 
\ln \left ( \frac{T_0}{\ep_0} \right ) \qquad ( s = 2) \; ,
$$
\be
\label{51}
 \frac{\Dlt T_c}{ T_0} = -\; \frac{\zeta(s-1)\ep_0}{s\zeta(s)T_0} \qquad
( s > 2 ) \;  .
\ee
Taking into account that $\varepsilon_0$ varies with increasing $N$ in the same way as
$\omega_0$, that is $\varepsilon_0 \propto N^{-1/s}$, one finds
$$
\frac{\Dlt T_c}{ T_0} \; \propto \; -\; \frac{1}{N^{1-1/s}} \qquad
( 1 < s < 2 ) \; ,
$$
$$
\frac{\Dlt T_c}{ T_0} \; \propto \; -\; \frac{\ln N}{\sqrt{N}} \qquad ( s= 2) \; ,
$$
\be
\label{52}
 \frac{\Dlt T_c}{ T_0} \; \propto \; -\; \frac{1}{N^{1/s}} \qquad
( s > 2 ) \;  .
\ee

As is seen, the corrections here are of different sign as compared to the previous 
section. They are of the same order of magnitude for $1 < s < 2$, but 
\be
\label{53}
\left |  \frac{\Dlt T_c}{ T_0} \right | \gg \frac{\Dlt T_0}{ T_0^{(0)}}  
\qquad ( s \geq 2 , \; N \ra \infty )
\ee
for large $N$. Remembering that the ideal Bose-condensed gas is stable only for $s > 2$,
we come to the conclusion that, for stable systems, the quantum corrections to the 
critical temperature, found in the present section, are more important than those of the 
previous section.  

For a three-dimensional harmonic trapping potential, one finds 
\cite{ Grossmann_66,Grossmann_67}. 
\be
\label{54}
 \frac{\Dlt T_c}{ T_0} = - \; 
\frac{\zeta(2)\overline\om}{2[\zeta(3)]^{2/3} \om_0 N^{1/3}} \qquad (s = d = 3) \;  ,
\ee
with the notation for the average trap frequency
$$
 \overline\om \equiv \frac{1}{d} \sum_{\al=1}^d \om_\al \;  .
$$
This can be rewritten as
\be
\label{55}
 \frac{\Dlt T_c}{ T_0} = - 0.727504 \left ( \frac{\overline\om}{\om_0} \right )
\frac{1}{N^{1/3}} \;  .
\ee
For anisotropic traps, $\bar{\omega} > \omega_0$. And in the case of a spherical 
harmonic trap, $\bar{\omega} = \omega_0$.

\section{Interaction corrections}

The semiclassical approximation can also be used for calculating the critical 
temperature shift under switching on weak interactions of atoms trapped in a 
harmonic potential. Above the critical temperature, the spectrum of atoms 
interacting through the local potential 
$$
 \Phi(\br) = \Phi_0\dlt(\br) \qquad 
\left ( \Phi_0 \equiv 4\pi \; \frac{a_s}{m} \right ) \;  ,
$$
in the semiclassical approximation, is
\be
\label{56}
 \ep_k(\br) = \frac{k^2}{2m} + U(\br) + 2\Phi_0\rho(\br) \qquad (T > T_c) \;  ,
\ee
where $\rho({\bf r})$ is the local density of atoms.  At the critical point, the chemical 
potential behaves as    
\be
\label{57}
 \mu \ra 2\Phi_0 \rho(0) \qquad (T \ra T_c + 0 ) \;  .
\ee

It is possible to show \cite{Yukalov_47} that the interaction corrections can
be small only for $s > 2$ and $d > 2$. Therefore in what follows, we take
$d = 3$ and assume that $s > 2$, which yields 
\be
\label{58}
 \frac{\Dlt T_c}{ T_0} = - c_3(s) \; \frac{a_s}{\lbd_0} \qquad ( s>2) \;  ,
\ee
with the thermal length
\be
\label{59}
 \lbd_0 \equiv \sqrt{\frac{2\pi}{mT_0} } = \sqrt{2\pi}\; [ \zeta(3) ]^{1/6}
\frac{l_0}{N^{1/6}} \;  ,
\ee
and with the notation
$$
 c_3(s) \equiv \frac{4}{s\zeta(s)} \left [ \zeta\left ( \frac{3}{2}\right ) \zeta(s-1)
- S_3(s) \right ] \; , \qquad
S_3(s) \equiv \sum_{m,n=1}^\infty \frac{n}{(mn)^{3/2}(m+n)^{s-3/2} } \;  .
$$
For a harmonic trap, when $s = d = 3$, one gets \cite{Pitaevskii_48}. 
\be
\label{60}
 \frac{\Dlt T_c}{ T_0} = - 3.426032 \; \frac{a_s}{\lbd_0} \;  .
\ee
This reproduces the linear term in expansion (\ref{17}). 

In the similar way, it is possible to find the critical temperature shift for other
types of interaction potentials, for instance for dipolar interactions
\cite{Glaum_68,Kao_69,Glaum_70}, for which the shift
$$
 \frac{\Dlt T_c}{ T_0} \; \propto \; \frac{a_D}{\lbd_0}  
$$
is proportional to the dipolar length $a_D \equiv m d_0^2$, with $d_0$ being 
dipolar moment.

Note that in the effective thermodynamic limit (\ref{34}) we have 
$\gamma_d \propto \omega_0^s$ and $\omega_0 \propto N^{- 1/s}$. For a 
three-dimensional trap $s = d = 3$, hence $\omega_0 \propto N^{- 1/3}$. 
Then $l_0 \propto N^{1/6}$, while $\lambda_0 \propto const$.

\section{Box-shaped trap}

Quasi-uniform traps of box shape have recently become available \cite{Gaunt_71}.
If the trap has the shape of a box of linear length $L$ and volume $V = L^3$, 
then the atomic wave function has to satisfy the boundary conditions
\be
\label{61}
 \psi(r_x,r_y,r_z) = 0 \qquad ( r_\al = 0,L) \;  .
\ee
The finite size of the box makes the condensation temperature $T_c$ different
from the transition temperature $T_0$ of an infinite homogeneous system. In the
case of ideal gas, 
\be
\label{62}
 T_0 = \frac{2\pi}{m} \left [ \frac{\rho}{\zeta(3/2)} \right ]^{2/3}
\qquad ( V \ra \infty ) \;  .
\ee
For the ideal gas, the shift of the critical temperature, caused by the finite size of 
the box, reads as \cite{Grossmann_72}
\be
\label{63}
 \frac{\Dlt T_c}{ T_0} = 0.351467 \; \frac{\ln N}{N^{1/3}} \;  .
\ee

However, if the box is strictly rectangular, and the Bose gas is ideal, then its 
condensation makes the system unstable, because of thermodynamically 
anomalous fluctuations, similarly to the homogeneous ideal Bose gas. This
follows \cite{Yukalov_47} from the number-of-particle variance
\be
\label{64}
{\rm var}(\hat N) = \frac{0.423}{2\pi^3} \; ( mT)^2 V^{4/3}
\ee
yielding the anomalous isothermal compressibility
\be
\label{65}
 \kappa_T = 0.00682 \; \frac{m^2 T}{\rho^{7/3}} \; N^{1/3} \;  .
\ee 

The ideal Bose-condensed gas is unstable either in an infinite homogeneous 
system or in a finite box-shaped trap. Fortunately, real atomic systems always
enjoy interactions that can stabilize the Bose-condensed gas.

\section{Optical lattices}

Cold atoms can be loaded in different optical lattices created by laser beams
\cite{Morsch_74,Moseley_75,Bloch_76,Yukalov_77}. The standard periodic 
potential, formed by the beams, reads as
\be
\label{66}
 V_L(\br) = \sum_{\al=1}^d V_\al \sin^2\left ( k_0^\al r_\al \right ) \;  ,
\ee
where ${\bf k_0}$ is the laser wave vector, 
$$
 k_0^\al \equiv \frac{2\pi}{\lbd_\al} = \frac{\pi}{a_\al}  ,
$$
$\lambda_\alpha$ is a laser wavelength, and $a_\alpha$ is a lattice spacing in 
the $\alpha$-direction. The lattice depth, or barrier, is characterized by the 
quantity
\be
\label{67}
 V_0 \equiv \frac{1}{d} \sum_{\al=1}^d V_\al \;  .
\ee
Atoms, subject to the action of the laser beams, have the recoil energy
\be
\label{68}
 E_R \equiv \frac{k_0^2}{2m} \qquad 
\left ( k_0^2 = \sum_{\al=1}^d (k_0^\al)^2 \right ) \;  .
\ee

The atomic field operator can be expanded over Wannier functions that can be
chosen to be well localized \cite{Marzari_78},
\be
\label{69}
\hat\psi(\br) = \sum_{jn} \hat c_{jn} w_n (\br - \ba_j ) \;   ,
\ee
here $j = 1,2, \ldots N_L$ enumerates the lattice sites and $n$ is a band index.
Substituting this expansion into the Hamiltonian, one usually considers only the 
lowest energy band.  Then, keeping in mind local interactions, characterized by the 
scattering length $a_s$, yields the Hubbard Hamiltonian
\be
\label{70}
 \hat H = - J \sum_{\lgl ij \rgl } \hat c_i^\dgr \hat c_j \; + \;
\frac{U}{2} \sum_j  \hat c_j^\dgr \hat c_j^\dgr \hat c_j \hat c_j \; + \; 
h_0 \sum_j  \hat c_j^\dgr \hat c_j   \;  .
\ee
The Hamiltonian parameters in three dimensions, resorting to the tight-binding 
approximation, can be presented  \cite{Yukalov_77} as follows. The tunneling 
parameter reads as
\be
\label{71}
  J = \frac{3}{4} \left ( \pi^2 - 4 \right ) V_0 \exp \left ( - \; 
\frac{3\pi^2}{4} \; \sqrt{ \frac{V_0}{E_R} } \right ) \;  .
\ee
The on-site interaction is described by 
\be
\label{72}
 U = \sqrt{ \frac{8}{\pi} } \; k_0 a_s E_R \left ( \frac{V_0}{E_R} \right )^{3/4} \; .
\ee
And the last term contains 
\be
\label{73}
  h_0 = 3 E_R \; \sqrt{ \frac{V_0}{E_R} } \; .
\ee
  
 In the presence of Bose-Einstein condensate, global gauge symmetry is broken,
which is the necessary and sufficient condition for the condensation
\cite{Lieb_79,Yukalov_80,Yukalov_81}. The most convenient way for breaking 
the symmetry is through the Bogolubov shift that here is equivalent to the canonical 
transformation
\be
\label{74}
 \hat c_j = \sqrt{\nu n_0} \; + \; c_j \;  ,
\ee
in which $\nu$ is the lattice filling factor 
\be
\label{75}
 \nu \equiv \frac{N}{N_L} = \rho a^d \;  ,
\ee
where $\rho$ is average atomic density, $a$ is a mean interatomic distance,
and $n_0 = N_0/N$ is the condensate fraction.

The operator of uncondensed atoms satisfies \cite{Yukalov_77} the properties
\be
\label{76} 
  \lgl c_j \rgl = 0 \; , \qquad \sum_j c_j = 0 \; .
\ee

In the Hartree-Fock-Bogolubov approximation, the critical temperature of
Bose-Einstein condensation, in $d$ dimensions, becomes \cite{Yukalov_77,Yukalov_82}
\be
\label{77}
 T_c =  4\pi \; \frac{d-2}{d} \left [ 
\Gm\left ( 1 + \frac{d}{2} \right ) \right ]^{2/d} J\nu \; .
\ee
In three dimensions, this reduces to
\be
\label{78}
 T_c =  5.064222 J \nu \;  .
\ee
In view of the tunneling parameter (\ref{71}), we have
\be
\label{79}
 \frac{T_c}{\nu E_R} = 22.294 \; \frac{V_0}{E_R} \; \exp \left ( - \; 
\frac{3\pi^2}{4} \; \sqrt{ \frac{V_0}{E_R} } \right ) \;  .
\ee
Recall that this expression is valid in the tight-binding approximation, when $E_R \ll V_0$. 

The dependence of the critical temperature on the lattice and interaction parameters was
studied by Monte Carlo simulations \cite{Nguyen_83}. The linear variation of $T_c$ with 
the filling factor, agreeing with Eq. (\ref{78}), is confirmed for $\nu < 6$.  

Particle fluctuations in the lattice are thermodynamically normal, with the atomic variance
$$
 {\rm var}(\hat N) = \frac{N T}{\nu U(n_0 +\sgm) } \; ,
$$
where
$$
\sgm = \frac{1}{N} \sum_j \lgl c_j c_j \rgl
$$
is the anomalous average. This gives the isothermal compressibility
\be
\label{80}
  \kappa_T = \frac{ {\rm var}(\hat N) }{\rho T N} =  
\frac{1}{\rho \nu ( n_0+\sgm) U } \;  .
\ee
As is seen, if the interaction parameter $U$ tends to zero, then the compressibility 
diverges, which means that the ideal Bose-condensed gas in a lattice is unstable,
while repulsive interactions stabilize it.      

Taking into account intersite atomic interactions results in the extended Hubbard 
model
$$
\hat H = - J \sum_{\lgl ij\rgl} \hat c_i^\dgr \hat c_j \; + \;
\frac{U}{2}  \sum_j  \hat c_j^\dgr \hat c_j^\dgr \hat c_j \hat c_j \; + \;
\frac{1}{2} \sum_{i\neq j} U_{ij} \hat c_i^\dgr \hat c_j^\dgr \hat c_i \hat c_j \; + \;
h_0 \sum_j \hat c_j^\dgr \hat c_j \; .
$$
This model is usually studied by means of numerical calculations, such as 
density-matrix renormalization group \cite{Kuhner_84,Kuhner_85} and quantum 
Monte Carlo simulations \cite{Niyaz_86,Batrouni_87,Wessel_88}. In the presence 
of intersite interactions collective phonon  excitations arise, which can make an
optical lattice unstable although long living \cite{Yukalov_89}.

\section{Conclusion}

The critical temperature of Bose-Einstein condensation is one of the main 
characteristics of Bose systems. In the present brief review, we give a survey 
of critical temperatures for typical systems experiencing Bose-Einstein condensation. 
These are weakly interacting uniform Bose gases, trapped Bose gases confined
in power-law potentials, and bosons in optical lattices. The emphasis is on the cases
allowing for the derivation of explicit expressions for the critical temperature. Such 
explicit expressions, even being approximate, make it possible to better understand
the dependence of the temperature on system parameters and to estimate optimal
conditions for realizing Bose-Einstein condensation.  

In the process of calculating a critical temperature, it is often necessary to employ 
elaborate mathematical methods. One such a very powerful method is based on
self-similar approximation theory. Since this approach can be useful for calculating
the critical temperature of different systems, it is sketched in the Appendix.    

\vskip 2mm

{\bf Appendix. Self-similar factor approximants}

\vskip 2mm

In many cases, the considered system is rather complicated allowing only for the use
of some kind of perturbation theory in powers of an asymptotically small parameter, 
while in real physical systems this parameter can be finite or even very large. 

Suppose we are interested in a function $f(x)$ of a real variable $x$, for which one 
can get an expansion in powers of this variable, obtaining
$$
f(x) \simeq f_k(x) \qquad ( x\ra 0 ) \;   ,
$$
with
$$
 f_k(x) =  f_0(x) \left ( 1 + \sum_{n=1}^k a_n x^n \right ) \;  ,
$$
where $f_0(x)$ is given. And assume that we need to know the value of the function
for finite, or even large, variable $x$. That is, we need a method of extrapolating
the function from asymptotically small $x$ to the region of finite $x$. A very powerful 
and rather simple method is based on self-similar approximation theory 
\cite{Yukalov_27,Yukalov_28,Yukalov_29,Yukalov_30,Yukalov_31,Yukalov_32,Yukalov_33}.
Here we briefly mention the main ideas of the theory and its particular approach of
extrapolation by self-similar factor approximants  \cite{Yukalov_34,Gluzman_35,Yukalov_36}.   

A finite series can be treated as a polynomial. By the fundamental theorem of algebra,
a polynomial 
$$
P_k(z) = \sum_{n=0}^k a_n z^n
$$
over the field of complex numbers (generally, with $a_n$ and $z$ complex-valued) can 
be uniquely presented as the product
$$
 P_k(z) = \prod_{j=1}^k ( a_j + b_j z) \;  .
$$
Therefore, we can write
$$
 \left ( 1 + \sum_{n=1}^k a_n x^n \right ) =  \prod_{j=1}^k ( 1 + b_j x) \; .
$$
The self-similar transformation of the linear function gives 
\cite{Yukalov_34,Gluzman_35,Yukalov_36}.  
$$
 ( 1 + b_j x) \ra ( 1 + A_j x)^{n_j} \;  .
$$
Then the series $f_k(x)$ transforms to the {\it self-similar factor approximant} 
$$
f_k^*(x) = f_0(x) \prod_{j=1}^{N_k} ( 1 + A_j x)^{n_j} \;   ,
$$
where $N_k$ will be given below and the parameters $A_j$ and $n_j$ are defined by the
accuracy-through-order procedure, that is, by equating the like-order terms in the 
small-variable expansion, so that 
$$
f_k^*(x) \simeq f_k(x) \qquad ( x\ra 0 ) \; .
$$

If the order $k$ is even, then $N_k = k/2$. And the accuracy-through-order procedure
yields $k$ equations
$$
 \sum_{j=1}^{k/2} n_j A_j^n = D_n \qquad ( n = 1,2,\ldots, k) \;  ,
$$
in which 
$$
D_n \equiv \frac{(-1)^{n-1}}{(n-1)!} \; \lim_{x\ra 0} \; \frac{d^n}{dx^n}
\; \ln \left ( 1 + \sum_{m=1}^n a_m x^m \right ) \;   .
$$
For instance,
$$
 D_1 = a_1 \; , \qquad D_2 = a_1^2 - 2a_2 \; , \qquad 
D_3 = a_1^3 - 3a_1a_2 + 3a_3 \;  .
$$
In these equations, there are $k/2$ unknown $A_j$ and $k/2$ unknown $n_j$, so that
the total number of unknowns equals the number of equations.    

If $k$ is odd, then we may set $N_k = (k + 1)/2$. However, then we again get $k$
equations
$$
 \sum_{j=1}^{(k+1)/2} n_j A_j^n = D_n \qquad ( n=1,2,\ldots, k ) \;  ,
$$
but now with $(k + 1)/2$ unknown $A_j$ and  $(k + 1)/2$ unknown $n_j$, which makes
$k + 1$ unknowns. One of the parameters remains undefined, requiring to impose
an additional condition. It is possible to resort to the scaling condition \cite{Yukalov_36},
agreeing to measure the parameters $A_j$ in units of one of them, say $A_1$. This is
equivalent to setting $A_1 = 1$. Thus
\begin{eqnarray}
\nonumber
N_k = \left \{
\begin{array}{ll}
k/2 \; , ~~& ~~ k = 2,4,\ldots \\
(k+1)/2 \; , ~~ & ~~ k = 3,5,\ldots
\end{array} \right. \; ,
\end{eqnarray} 
with the scaling condition $A_1 = 1$ for odd $k$. 

The accuracy-through-order equations with respect to $A_j$ are polynomial, the first
equation being of first order, the second, of second order, and so on, with the last 
equation being of order $k$. Each polynomial equation of order $n$ possesses $n$
solutions. So that the total number of solutions is $1 \times 2 \times 3 \cdots \times k = k!$.
At the same time, from the form of the factor approximants it is evident that each of them
is invariant with respect to the $k!$ permutations 
$$
 A_i \ra A_j \; , \qquad n_i \ra n_j \;  .
$$
Therefore, the multiplicity of solutions for $A_j$ is trivial, being related to the 
enumeration of the parameters. Up to this enumeration, the solutions are unique.  

If the parameters $A_j$ are given, then we get the linear algebraic equations with respect 
to $n_j$. The solutions for the latter have the form
$$
 n_j = \frac{U_j(N_k)}{V(N_k)} \qquad ( j = 1,2,\ldots, N_k) \;  .
$$
Here the nominator is the determinant
\begin{eqnarray}
\nonumber
U_j(N_k) = \left | \begin{array}{cccccc}
A_1 & A_2 & \ldots & D_1 & \ldots & A_{N_k} \\
A_1^2 & A_2^2 & \ldots & D_2 & \ldots & A_{N_k}^2 \\
\ldots & \ldots& \ldots & \ldots & \ldots & \ldots \\
A_1^{N_k} & A_2^{N_k} & \ldots & D_{N_k} & \ldots & A_{N_k}^{N_k} 
\end{array} \right | \; ,
\end{eqnarray}
with $D_i$ in the $j$-th column, and the denominator is the Vandermonde-Knuth 
determinant
\begin{eqnarray}
\nonumber
V(N_k) = \left | \begin{array}{cccccc}
A_1 & A_2 & \ldots & A_j & \ldots & A_{N_k} \\
A_1^2 & A_2^2 & \ldots & A_j^2 & \ldots & A_{N_k}^2 \\
\ldots & \ldots& \ldots & \ldots & \ldots & \ldots \\
A_1^{N_k} & A_2^{N_k} & \ldots & A_j^{N_k} & \ldots & A_{N_k}^{N_k} 
\end{array} \right | \; .
\end{eqnarray}
When $k$ is odd, hence $A_1 = 1$, then $V(N_k)$ is the standard Vandermonde 
determinant, while when $k$ is even, this determinant by the relation
$$
V(N_k) = \left ( \prod_{j=1}^{N_k} A_j \right ) \overline V(N_k)
$$
is connected with the Vandermonde determinant
\begin{eqnarray}
\nonumber
\overline V(N_k) = \left | \begin{array}{cccccc}
1   &  1  & \ldots & 1   & \ldots & 1 \\
A_1 & A_2 & \ldots & A_j & \ldots & A_{N_k} \\
A_1^2 & A_2^2 & \ldots & A_j^2 & \ldots & A_{N_k}^2 \\
\ldots & \ldots& \ldots & \ldots & \ldots & \ldots \\
A_1^{N_k-1} & A_2^{N_k-1} & \ldots & A_j^{N_k-1} & \ldots & A_{N_k}^{N_k-1} 
\end{array} \right | =  \prod_{1\leq i < j \leq N_k} (A_j - A_i ) \; .
\end{eqnarray}
The beauty of the factor approximants is that they, in a finite order, can exactly 
reconstruct a large class of functions from their asymptotic expansions. This class 
includes rational, irrational, as well as transcendental functions  
\cite{Yukalov_34,Gluzman_35,Yukalov_36}. For example, exactly reproducible 
are all functions of the type
$$
R(x) = \prod_{j=1}^M P_{m_j}^{\al_j}(x) \;   ,
$$
where $P_m(x)$ are polynomials and $\alpha_j$ are complex-valued numbers. To 
exactly reconstruct this function, one needs a factor approximant of the order
$$
 k = \sum_{j=1}^M m_j \;  .
$$
 
Exactly reproducible are transcendental functions that can be defined as limits
of polynomials \cite{Yukalov_36}. For instance, as is easy to check, since
$$
\lim_{a\ra 0} P_1^{1/a}(x) = e^x \qquad ( P_1(x) = 1 + ax ) \;   ,
$$
the exponential function is exactly reproduced in any order $k \geq 2$.  Keeping 
in mind such limits, the class of exactly reproducible functions can be denoted as
$$
{\cal R} = \{ R(x) , \; \lim R(x) \} \;   .
$$

Self-similar factor approximants extrapolate asymptotic series to finite values of
the variables, and even to the variables tending to infinity. Thus, assuming that
$$
 f_0(x) \simeq A x^\al \qquad ( x\ra \infty ) \;  ,
$$
we get the large-variable behavior of the factor approximant as
$$
 f_k^*(x) \simeq B_k x^\bt_k \qquad ( x\ra \infty ) \;  ,
$$
with the amplitude
$$
B_k =  A \prod_{j=1}^{N_k} A_j^{n_j}
$$
and the power
$$
 \bt_k = \al + \sum_{j=1}^{N_k} n_j \;  .
$$

When the large-variable behavior of the sought function is known, say
$$
 f(x) \simeq B x^\bt \qquad (x \ra \infty) \;  ,
$$
then it is possible to require that $\beta_k$ be equal to $\beta$, 
$$
\al + \sum_{j=1}^{N_k} n_j = \bt \; .
$$
If it is known that the limit of the sought function is finite, hence $\beta = 0$, then 
one has the condition
$$
\al + \sum_{j=1}^{N_k} n_j = 0 \qquad ( \bt = 0 ) \; .
$$

Thus, for a finite series $f_k(x)$, one gets a sequence $\{ f_n^*(x)\}$ of self-similar 
factor approximants, with $n = 1, 2, \ldots k$. If the sequence converges numerically, 
then as the final answer one can accept the expression $[f_k^*(x) + f_{k-1}^*(x)] / 2$, 
with the error bar  $\pm \vert f_k^*(x) - f_{k-1}^*(x) \vert /2$. 

The convergence of the sequence $\{ f_n^*(x)\}$ can be accelerated in the following way.  
Being based on the last three factor approximants, one constructs a quadratic spline
$$
 q(x,t) = a(x) + b(x) t + c(x) t^2 \;   ,
$$
whose coefficients are determined from the conditions
$$
q(x,0) = f_{k-2}^*(x) \; , \qquad q(x,1) = f_{k-1}^*(x) \; , \qquad
q(x,2) = f_k^*(x) \;   .
$$
The factor approximant for the spline is
$$
q^*(x,t) = a(x) [ 1 + A(x) t]^{n(x)} \;   .
$$
As the final answer, one can accept 
$$
 q^*(x) = \frac{1}{2} \left [ q^*(x,2) + q^*(x,3) \right ] \;   ,
$$
with the error bar $\pm \vert q^*(x,2) - q^*(x,3) \vert /2$. 

The found $q^*(x)$ extrapolates the initial asymptotic series to finite values
of the variable $x$, including the limit $x \ra \infty$.  

This method was used \cite{Yukalov_73} for calculating the limit (\ref{7}) 
characterizing the critical-temperature shift for the uniform Bose gas in Sec. 2. 
The results for $c_1$ are listed in Table 2.

\newpage

\begin{table}
\centering
\label{tab-1}
\renewcommand{\arraystretch}{1.75}
\begin{tabular}{|c|c|c|c|c|c|} \hline
$N$   &         0    &       1      &    2          &    3    &    4     \\ \hline
$a_1$ &    0.111643  &    0.111643  &    0.111643   &    0.111643   &    0.111643 \\ \hline
$a_2$ & $-$0.0264412 & $-$0.0198309 & $-$0.0165258  & $-$0.0145427  & $-$0.0132206 \\ \hline
$a_3$ &    0.0086215 &    0.00480687&    0.00330574 &    0.00253504 &    0.0020754 \\ \hline
$a_4$ & $-$0.0034786 & $-$0.00143209& $-$0.000807353& $-$0.000536123& $-$0.000392939 \\ \hline
$a_5$ &    0.00164029&    0.00049561&    0.000227835&    0.000130398&    0.0000852025 \\ \hline
\end{tabular}
\vskip 3mm
\caption{Coefficients $a_n$ of the asymptotic expansion for $c_1(x)$,
calculated in seven-loop perturbation theory \cite{Kastening_26} for different
number of components $N$.}
\end{table}

\newpage

\begin{table}
\centering
\label{tab-2}
\renewcommand{\arraystretch}{1.75}
\begin{tabular}{|c|c|c|} \hline
$N$ &     $c_1$      &       Monte Carlo                       \\ \hline
0   & 0.77$\pm$ 0.03 &                                         \\ \hline
1   & 1.06$\pm$ 0.05 & 1.09$\pm$ 0.09  \cite{Sun_37}           \\ \hline
2   & 1.29$\pm$ 0.07 & 1.29$\pm$ 0.05 \cite{Kashurnikov_6}    \\ 
    &                & 1.32$\pm$ 0.02 \cite{Arnold_8}  \\ \hline
3   & 1.46$\pm$ 0.08 &                       \\ \hline
4   & 1.60$\pm$ 0.09 & 1.60$\pm$ 0.10 \cite{Sun_37}              \\ \hline
\end{tabular}
\vskip 3mm
\caption{Coefficient $c_1$ of the critical temperature, for different number of 
components $N$, found by using self-similar factor approximants compared with the 
available Monte Carlo simulations.}
\end{table}

\newpage

\begin{table}
\centering
\label{tab-3}
\renewcommand{\arraystretch}{1.75}
\begin{tabular}{|c|c|c|c|c|c|} \hline
$N$   &         0    &       1      &    2          &    3    &    4     \\ \hline
$b_0$ &    4412.37   &    6618.56   &    8824.74    &    11030.9    &    13237.1 \\ \hline
$b_1$ &    58.5209   &    87.7814   &    117.042    &    146.302    &    175.563 \\ \hline
$b_2$ & $-$2.393297  & $-$2.393297  &  $-$2.393297  & $-$2.393297   & $-$2.393297 \\ \hline
$b_3$ & $-1.96607\cdot 10^{-2}$ & $-1.43753\cdot 10^{-2}$ & $-1.17327\cdot 10^{-2}$ & $-1.01470\cdot 10^{-2}$ & $-9.08997\cdot 10^{-3}$ \\ \hline
$b_4$ & $3.30862\cdot 10^{-4}$  & $1.78649\cdot 10^{-4}$  & $1.19596\cdot 10^{-4}$  & $8.96215\cdot 10^{-5}$  & $7.19120\cdot 10^{-5}$ \\ \hline
$b_5$ & $-8.82744\cdot 10^{-6}$ & $-3.51348\cdot 10^{-6}$ & $-1.92434\cdot 10^{-6}$ & $-1.24601\cdot 10^{-6}$ & $-8.93033\cdot 10^{-7}$ \\ \hline
$b_6$ & $2.99799\cdot 10^{-7}$  & $8.75642\cdot 10^{-8}$  & $3.90859\cdot 10^{-8}$  & $2.17953\cdot 10^{-8}$  & $1.39127\cdot 10^{-8}$ \\ \hline
$b_7$ & $-1.19671\cdot 10^{-8}$  & $-2.55154\cdot 10^{-9}$ & $-9.24418\cdot 10^{-10}$& $-4.42457\cdot 10^{-10}$& $-2.5085\cdot 10^{-10}$ \\ \hline
\end{tabular}
\vskip 3mm
\caption{Coefficients $b_n$ of the asymptotic expansion for $d_N(x)$,
 calculated in seven-loop perturbation theory \cite{Kastening_26} for different number 
of components $N$.}
\end{table}

\newpage

\begin{table}
\centering
\label{tab-4}
\renewcommand{\arraystretch}{1.75}
\begin{tabular}{|c|c|c|} \hline
$N$ &     $d_N$  (Monte Carlo)    &   $d_N$ (OPT) \cite{Kastening_26}   \\ \hline
0   &                                      &  0.636 $\pm$ 0.01        \\ \hline                            
1   & 0.898 $\pm$ 0.004 \cite{Sun_37} &  0.893 $\pm$ 0.01        \\ \hline
2   & 1.059 $\pm$ 0.001 \cite{Arnold_9}    &  1.060 $\pm$ 0.01        \\ \hline
3   &                                      &  1.178 $\pm$ 0.01        \\ \hline
4   & 1.255 $\pm$ 0.006 \cite{Sun_37}  &  1.265 $\pm$ 0.01        \\ \hline
\end{tabular}
\vskip 3mm
\caption{Parameters $d_N$ entering the coefficients $c_2$, found by Monte Carlo 
simulations and optimized perturbation theory, for different number of components $N$.}
\end{table}

\begin{table}
\centering
\label{tab-5}
    \renewcommand{\arraystretch}{1.75}
\begin{tabular}{|c|c|c|c|c|c|c|c|c|c|} \hline
$\gm$     & 0.00464 & 0.00794 & 0.01   & 0.0215 & 0.0464 & 0.126  & 0.171  & 0.215  & 0.368 \\ \hline
$T_c/T_0$ & 1.0069  & 1.0091  & 1.0127 & 1.0214 & 1.0351 & 1.0624 & 1.0652 & 1.0627 & 1.0060 \\ \hline
\end{tabular}
\vskip 3mm
\caption{Relative critical temperature $T_c/T_0$ as a function of the gas
parameter $\gm$, obtained by Monte Carlo simulations \cite{Pilati_39} for three-dimensional
homogeneous Bose system.}
\end{table}

\newpage

\begin{figure}[ht!]
\vspace{9pt}
\centerline{
\includegraphics[width=10cm]{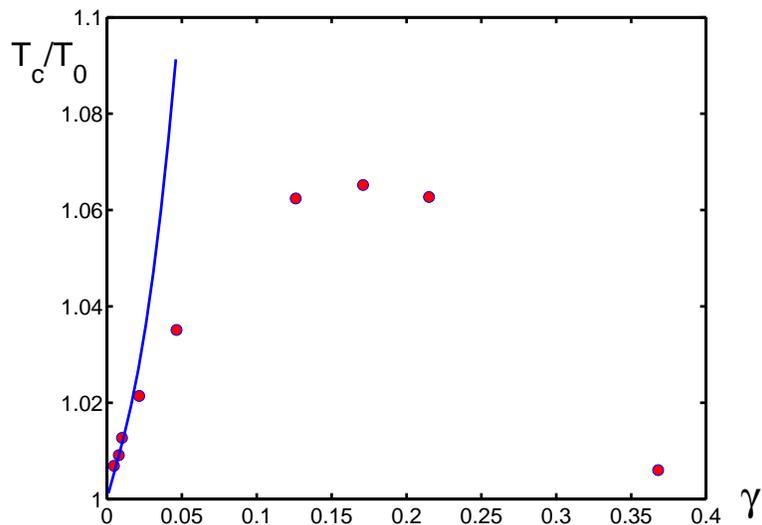} }
\caption{Relative critical temperature as a function of the gas parameter
for a homogeneous system with $N=2$. Solid line corresponds to the asymptotic
expression (\ref{4}); dots are the Monte Carlo results from Table 5.
}
\label{fig:Fig.1}
\end{figure}

\begin{figure}[ht!]
\vspace{9pt}
\centerline{
\includegraphics[width=10cm]{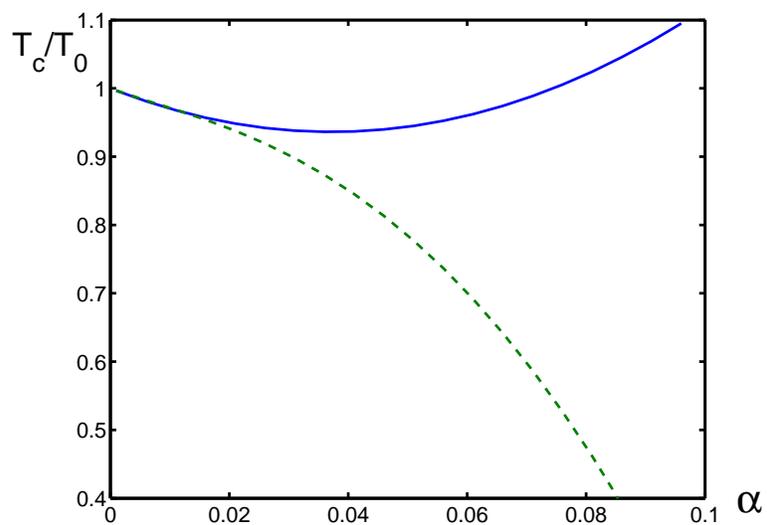} }
\caption{Relative critical temperature of Bose-Einstein condensation of trapped
atoms, corresponding to Eq. (\ref{17}) (dashed line) and to Eq. (\ref{20}) (solid
line), as a function of the coupling parameter $\al$.
}
\label{fig:Fig.2}
\end{figure}
\end{document}